\documentclass{PoS}

\usepackage{amssymb, amsbsy, amsfonts}
\usepackage{amsmath, amsthm, cite}

\usepackage{float}

\newcommand\N{\nonumber}

\newcommand\HA{{\rm H}}
\newcommand\LL{{\rm L}}

\newcommand{\gsim}{\raisebox{-0.07cm   }
{$\, \stackrel{>}{{\scriptstyle\sim}}\, $}}

\allowdisplaybreaks[3]

{

\newcounter{mmacnt}
\def\restartmma{\setcounter{mmacnt}{0}}
\restartmma \catcode`|=\active
\def|#1|{\mathrm{#1}}
\catcode`|=12

\title{{\footnotesize DESY 12-232, DO-TH-12/38, SFB/CPP-12-106, LPN12-142 \hfill
{\tt arXive:1212.6823[hep-ph]}
}
\\
Three-Loop Contributions to the Gluonic Massive Operator Matrix Elements at General 
Values of N}
\ShortTitle{Three-Loop Contributions to the Gluonic Massive OMEs}

\author{Jakob Ablinger$^{a,b}$, \speaker{Johannes Bl\"umlein}$^b$, 
        Abilio De Freitas$^a$, Alexander
        Hassel- huhn$^{a,b}$, Sebastian Klein$^c$, Clemens Raab$^b$, Mark Round$^a$, 
        Carsten Schneider$^a$, Fabian Wi\ss{}brock$^a$\\
        $^a$Research Institute for Symbolic Computation (RISC),
        Johannes Kepler University Linz,
        Altenberger Str. 69, A-4040 Linz, Austria\\
        $^b$Deutsches Elektronen--Synchrotron (DESY), Zeuthen     
        Platanenallee 6, D-15738 Zeuthen, Germany \\
        $^c$ Institut f\"ur Theoretische Physik E, RWTH Aachen University, D-52056 
        Aachen, Germany}

\abstract{\noindent
Recent results on the calculation of 3-loop massive operator matrix elements in case of 
one and two heavy quark masses are reported. They concern the $O(n_f T_F^2 C_{F,A})$ 
and $O(T_F^2 C_{F,A})$ gluonic corrections, two-mass quarkonic moments, and ladder- and 
Benz-topologies. We also discuss technical aspects of the calculations.}

\FullConference{Loops and Legs in Quantum Field Theory - 11th DESY Workshop on Elementary Particle Physics,\\
		April 15-20, 2012\\
		Wernigerode, Germany}

\begin{document}

%%%%%%%%%%%%%%%%%%%%%%%%%%%%%%%%%%%%%%%%%%%%%%%%%%%%%%%%%%%%%%%%%%%%%%%%%%%%%%%%%%%%%%%%%%
\section{Introduction}
%%%%%%%%%%%%%%%%%%%%%%%%%%%%%%%%%%%%%%%%%%%%%%%%%%%%%%%%%%%%%%%%%%%%%%%%%%%%%%%%%%%%%%%%%%

\vspace{1mm}
\noindent
One of the most precise determinations of the strong coupling constant $\alpha_s(M_Z)$
relies on the next-to-next-to-leading order (NNLO) QCD analysis of the world 
deep-inelastic data \cite{Alekhin:2012ig,Bethke:2011tr}. Here currently the heavy 
flavor corrections to the structure function $F_2(x,Q^2)$ and transversity are known 
for the first Mellin moments $N = 2,..,10(14)$ 
\cite{Bierenbaum:2009mv,Blumlein:2009rg}.\footnote{Present analyses use the NLO corrections 
in $x$-space \cite{laenen1} or Mellin space implementations \cite{Alekhin:2003ev}.} 
In the asymptotic 
region $Q^2 \gg m^2$ the heavy flavor Wilson coefficients can be represented in form of
convolutions \cite{Buza:1995ie} of massive operator matrix elements (OMEs) and 
the massless Wilson coefficients \cite{Vermaseren:2005qc}. In case of the charm quark 
contribution the corresponding region is given by $Q^2/m_c^2 \gsim 10$. To carry out
complete NNLO QCD analyses in this region the heavy flavor Wilson coefficients have 
to be known for general values of $N$. This also applies to precision measurements of
the charm quark mass using the world deep-inelastic data  \cite{Alekhin:2012vu}.

Since 2010 the 
systematic calculation of the asymptotic massive Wilson coefficients at general 
values of $N$ have been carried out. Five massive Wilson coefficients contribute to 
$F_2(x,Q^2)$ at NNLO \cite{Bierenbaum:2009mv}. Furthermore, there are other massive 
OMEs needed to compute the matching coefficients in the variable flavor number scheme
in which the heavy quarks are assumed to decouple singly 
\cite{Buza:1996wv,Bierenbaum:2009zt,Bierenbaum:2009mv}. There are yet other 
contributions at NNLO from graphs containing both a massive charm and a bottom quark 
line, which extend the former representation in Ref.~\cite{Bierenbaum:2009mv} and are 
necessary because of the fact that charm quarks do not yet become massless at the mass 
scale 
of bottom quarks since $m_c^2/m_b^2 \approx 1/10$ only. Results on first moments
for these contributions are obtained in \cite{Ablinger:2011pb,Ablinger:2012qj,BW13}.
The extension of the renormalization of the massive OMEs is given in \cite{BW13}.
The logarithmic contributions at general values of $N$ for the contributions to 
$F_2(x,Q^2)$ are available \cite{Bierenbaum:2010jp,BBKW13,Kawamura:2012cr}.\footnote{
The $O(\alpha^2_s)$ and $O(\alpha^2_s \varepsilon)$ contributions, with  $\varepsilon = 
D-4$ the dimensional parameter, needed in the  renormalization were given in 
\cite{Buza:1995ie,Bierenbaum:2007qe,Bierenbaum:2008yu}.}  
Furthermore, the NNLO heavy flavor Wilson coefficients in the asymptotic region were 
calculated for the structure function $F_L(x,Q^2)$ \cite{Blumlein:2006mh,Bierenbaum:2009mv}.

Two of the five massive Wilson coefficients contributing to $F_2(x,Q^2)$ at NNLO are
known completely \cite{Ablinger:2010ty}. They are of $O(n_f T_F^2 C_{F,A})$. Likewise,
these contributions to the further three  Wilson coefficients and transversity were 
calculated for these color coefficients in \cite{Ablinger:2010ty}. Also the complete 
contribution $O(T_F^2 C_{A,F})$ for the OMEs $A_{Qq}^{\rm PS}, A_{qq,Q}^{\rm NS}, 
A_{qq,Q}^{\rm TR}$ are available \cite{Ablinger:2011pb}.

In these proceedings we report on the calculation of the gluonic OMEs $O(n_f T_F^2 
C_{F,A})$ \cite{Blumlein:2012vq} in Section~2 and on first results in $O(T_F^2 C_{F,A})$ 
for this channel in Section~3. Furthermore, the scalar 3-loop integrals for all ladder 
type integrals were calculated \cite{Ablinger:2012qm,ABSW13}. An extension of the method 
\cite{Brown:2008um} to calculate finite Feynman integrals to the case of massive quark 
lines with local operator insertions has been used to calculate ladder- and 
Benz-topologies \cite{Ablinger:2012qm,ABSW13}, also leading to new types of finite 
nested sums extending the harmonic \cite{HSUM}, generalized harmonic 
\cite{Moch:2001zr,GENS2}, and cyclotomic sums \cite{Ablinger:2011te} and the 
associated polylogarithms, cf. Section~4.
Section~5 contains the conclusions.
   
%%%%%%%%%%%%%%%%%%%%%%%%%%%%%%%%%%%%%%%%%%%%%%%%%%%%%%%%%%%%%%%%%%%%%%%%%%%%%%%%%%%%%%%%%%
\section{\bf Gluonic OMEs $\mathbf{O(n_f T_F^2 C_{F,A})}$}
%%%%%%%%%%%%%%%%%%%%%%%%%%%%%%%%%%%%%%%%%%%%%%%%%%%%%%%%%%%%%%%%%%%%%%%%%%%%%%%%%%%%%%%%%%

\vspace{1mm}
\noindent
The calculation of all $O(n_f T_F^2 C_{F,A})$ contributions to the massive OMEs has 
been completed with the computation of $A_{gg,Q}$ and $A_{gq,Q}$ at this order in 
\cite{Blumlein:2012vq}.
In these and other computations described below we used the codes 
{\tt QGRAF} \cite{Nogueira:1991ex}, {\tt Form} and {\tt tform} \cite{FORM}, and {\tt 
color} \cite{vanRitbergen:1998pn}. For the check of individual moments of the 
expressions derived we also used {\tt MATAD} \cite{Steinhauser:2000ry}.

In the calculation of the $O(n_f T_F^2 C_{F,A})$-terms large amounts of nested sums 
emerge. An important step in the 
calculation consists in merging these individual sums to a smaller amount of sums
containing very voluminous summands which are then solved with the codes used within the
package {\tt Sigma} \cite{sigma}, like {\tt EvaluateMultiSum} and {\tt SumProduction}, 
\cite{CSchneider1}, written in {\tt mathematica} \cite{MATHEM}, cf. also \cite{TECHN}. 
Moreover, this 
compactification shall be performed for whole diagrams to avoid the intermediate emergence 
of a larger amount of generalized harmonic sums as was observed in \cite{Ablinger:2010ty}.
Generalized harmonic sums do not contribute in this case.

The constant contributions to the unrenormalized massive OMEs $A_{gg,Q}$ and 
$A_{gq,Q}$, $a_{gj,Q}^{(3)},~j = q,g$, 
read~:
%--------------------------------------------------------------------
\begin{eqnarray}
  a_{gq,Q}^{(3),n_fT_F^2}
  &=& C_F T_F^2 n_f \Biggl\{ 
  -\frac{16 \left(N^2+N+2\right)}{9  (N-1) N (N+1)}
   \left( \frac{1}{3} S_{1}^3 + S_{2} S_{1} + \frac{2}{3} S_{3} +
         14 \zeta_3 + 3 S_{1} \zeta_2 \right)
\N\\ & &
  +\frac{16 \left(8 N^3+13 N^2+27 N+16\right)}{27 (N-1) N (N+1)^2} 
   \left( 3 \zeta_2 + S_{1}^2 + S_{2}\right)
\N\\ & &
  -\frac{32 \left(35 N^4+97 N^3+178 N^2+180 N+70\right)}{27 (N-1) N
(N+1)^3} S_{1}
\N\\ & &
  +\frac{32 \left(1138 N^5+4237 N^4+8861 N^3+11668 N^2+8236
N+2276\right)}{243 (N-1) N (N+1)^4}
 \Biggr\}~,
\\
%\end{eqnarray}
%--------------------------------------------------------------------
%--------------------------------------------------------------------
%\begin{eqnarray}
  a_{gg,Q}^{(3),n_f T_F^2}
  &=& n_f T_F^2 \Biggl\{
  C_A 
  \frac{1}{(N-1) (N+2)}
  \Biggl[
    \frac{4 P_{1}}{27 N^2 (N+1)^2} S_{1}^2
   +\frac{8 P_{2}}{729 N^3 (N+1)^3} S_{1}
\N\\
&&   +\frac{160}{27} (N-1) (N+2) \zeta_2 S_{1}
   -\frac{448}{27} (N-1) (N+2) \zeta_3 S_{1}
   +\frac{P_{3}}{729 N^4 (N+1)^4}
\N\\
&&
   -\frac{2 P_{4}}{27 N^2 (N+1)^2} \zeta_2
   +\frac{56 \left(3 N^4+6 N^3+13 N^2+10 N+16\right)}{27 N (N+1)} \zeta_3
   -\frac{4 P_{5}}{27 N^2 (N+1)^2} S_{2}
  \Biggr]
\N\\&&
  +C_F 
   \frac{1}{(N-1) (N+2)}
   \Biggl[
     \frac{112 \left(N^2+N+2\right)^2}{27 N^2 (N+1)^2} S_{1}^3
    -\frac{16 P_{6}}{27 N^3 (N+1)^3} S_{1}^2
\N\\&&
    +\frac{32 P_{7}}{81 N^4 (N+1)^4} S_{1}
    +\frac{16 \left(N^2+N+2\right)^2}{3 N^2 (N+1)^2} \zeta_2 S_{1}
    +\frac{16 \left(N^2+N+2\right)^2}{3 N^2 (N+1)^2} S_{2} S_{1}
\N\\&&
    -\frac{32 P_{8}}{243 N^5 (N+1)^5}
    -\frac{16 P_{9}}{9 N^3 (N+1)^3} \zeta_2
    +\frac{448 \left(N^2+N+2\right)^2}{9 N^2 (N+1)^2} \zeta_3
    +\frac{16 P_{10}}{9 N^3 (N+1)^3} S_{2}
\N\\&&
    -\frac{160 \left(N^2+N+2\right)^2}{27 N^2 (N+1)^2} S_{3}
   \Biggr]\Biggr\}~.
\end{eqnarray}
%----------------------------------------------------------------------------------
The corresponding $1/\varepsilon$ terms contain contributions of the
3-loop anomalous dimensions \cite{Vogt:2004mw} which we have verified by an 
independent calculation. Moreover a prediction made in \cite{Bennett:1997ch} 
has been confirmed.

The gluonic OMEs  $A_{gg,Q}$ and $A_{gq,Q}$ are needed for correct flavor matching
in case of the transition of a {\it single} heavy quark becoming light, 
cf.~\cite{Buza:1996wv,Bierenbaum:2009mv}. Here the correct choice of the matching scale 
is of importance \cite{Blumlein:1998sh}.

In extending the above calculation the computation of some topological classes of diagrams 
has been automated mapping the graph to expressions involving hypergeometric 
$_{p+1}F_p$-functions. The $\varepsilon$-expansion leads to nested sums being calculated 
using the package {\tt Sigma} \cite{sigma}. The automation will cover other classes soon, 
requiring more involved ways to match the initial functions.
%%%%%%%%%%%%%%%%%%%%%%%%%%%%%%%%%%%%%%%%%%%%%%%%%%%%%%%%%%%%%%%%%%%%%%%%%%%%%%%%%%%%%%%%%%
\section{Gluonic OMEs $\bf{O(T_F^2 C_{F,A})}$ and
OMEs with massive fermion lines of two different masses}
%%%%%%%%%%%%%%%%%%%%%%%%%%%%%%%%%%%%%%%%%%%%%%%%%%%%%%%%%%%%%%%%%%%%%%%%%%%%%%%%%%%%%%%%%%

\vspace{1mm}
\noindent
All basic scalar topologies contributing to the  $O(T_F^2 C_{F,A})$ of $A_{gg,Q}$ and 
$A_{gq,Q}$ have been calculated. An example is given in (\ref{eq:TF2}) for the 
graph containing two massive triangles with $m_1 = m_2$,
%---------------------------------------------------------------------------------
\begin{eqnarray}
I_{D2}(N,\varepsilon) &=& (-1)^N \Biggl\{
-\frac{1
%(-1)^N
}{12 N \varepsilon}+
\frac{%(-1)^N 
\big(27 N^2-5 N+16\big)}{1440 N (N+1) (2 N-1) 4^N} \binom{2N}{N} 
\left[
\sum_{\text{i}_1=1}^N \frac{4^{i_1}}{i_1^2} \frac{ 
S_1\big({i}_1-1\big)}{\displaystyle \binom{2i_1}{i_1}} - 7 \zeta_3\right] 
%\big(\text{i}_1!\big){}^2 S_1\big(\text{i}_1-1\big)}{\big(2 \text{i}_1\big)! 
%\text{i}_1^2} 
%- \sum_{\text{i}_1=1}^N \frac{2^{2 \text{i}_1} \big(
%\text{i}_1!\big){}^2}{\big(2 \text{i}_1\big)! \text{i}_1^3}
%\right)
%\N\\
%&&+\frac{(-1)^N 2^{-2 N-5} (2 N)!}{N (N+1) 
%(2 N-1) (N!)^2} \frac{1}{45} \big(-27 N^2+5 N-16\big) 
%\sum_{\text{i}_1=1}^N \frac{2^{2 \text{i}_1} \big(
%\text{i}_1!\big){}^2}{\big(2 \text{i}_1\big)! \text{i}_1^3}
%\N\\
%&&
%+(-1)^N 
%\zeta_3 \big(
%\frac{7}{90 N}
%-\frac{7}{45} \big(27 N^2-5 N+16\big) 
%\frac{2^{-2 N-5} (2 N)!}{N (N+1) (2 N-1) (N!)^2}\big)
%
\N\\
&&
-\frac{%(-1)^N 
\left(S_{2,1}-S_3 - 7 \zeta_3\right)}{90 N}-\frac{%(-1)^N 
\left(S_1^2-S_2\right)}{90 
(N-1) N^2 
(N+1)}
%+\frac{%(-1)^N 
%S_3(N)}{90 N} 
%+\frac{%(-1)^N 
%S_2(N)}{90 (N-1) N^2 (N+1)}
\N\\ &&
+\frac{
%(-1)^N 
\big(60 N^3-19 N^2-85 N+60\big)}{720 (N-1) N (N+1) (2 N-1)} S_1
+\frac{%(-1)^N 
\big(-162 N^3+281 N^2-187 N+30\big)}{720 (N-1) N^2 (2 N-1)}
\Biggr\}~.
\label{eq:TF2}
\end{eqnarray}
%---------------------------------------------------------------------------------
Here we chose a minimal representation of sums being pairwise transcendent,
which is proven by {\tt Sigma} \cite{sigma}.
In these and similar diagrams sums of the kind 
%---------------------------------------------------------------------------------
\begin{eqnarray}
\label{eq:NSU1}
\sum_{i = 1}^N \frac{4^i}{i^3}\frac{1}{\displaystyle \binom{2i}{i}} &=& \int_0^1 dx
\frac{x^N-1}{x-1} \HA^*_{\sf 0 w_3}(x)
\\
%-------
\label{eq:NSU2}
\sum_{i = 1}^N \frac{4^i}{i^2}\frac{1}{\displaystyle \binom{2i}{i}} S_1(i) &=&
\int_0^1 dx \frac{x^N-1}{x-1}\left[\HA^*_{\sf 0 w_3}(x)-\HA^*_{\sf w_3 0}(x)
-\HA^*_{\sf w_3 1}(x)-2\ln(2)\HA^*_{\sf w_3}(x)\right]
\end{eqnarray}
%----------------------------------------------------------------------------
emerge, which can be represented as Mellin transforms of iterated integrals, 
extending the usual harmonic polylogarithms 
(HPLs) \cite{Remiddi:1999ew}, where $\HA^*_\emptyset(x) = 1, 
\HA^*_{\sf b,\vec{c}}(x) = \int_x^1 dy f_{\sf b}(y) \HA^*_{\sf \vec{c}}(y), 
H^*_1(x) = \ln(1-x)$, 
and $f_{\sf w_3}(x) = 1/(x\sqrt{1-x})$, cf.~\cite{ABRS13}. 
The relative transcendence of the respective
HPLs is proven using differential field methods \cite{BRON}, cf. also~\cite{RAAB}. 
In exceptional cases they may be obtained in terms of HPLs with root
arguments. Eq. (\ref{eq:TF2}) obviously is recurrent in $N$. The asymptotic 
expansions of (\ref{eq:NSU1}, \ref{eq:NSU2}) are given by
%---------------------------------------------------------------------------------
\begin{eqnarray}
\sum_{i = 1}^N \frac{4^i}{i^3}\frac{1}{\displaystyle \binom{2i}{i}}
&=&
6 \zeta_2 \ln(2) - \frac{7}{2} \zeta_3 - \sqrt{\frac{\pi}{N}}\Biggl\{
\frac{2}{3} \frac{1}{N} - \frac{9}{20} \frac{1}{N^2} + \frac{199}{1344} \frac{1}{N^3}
+ O\left(\frac{1}{N^4}\right) \Biggr\}\\
%-------------
\sum_{i = 1}^N \frac{4^i}{i^2}\frac{1}{\displaystyle \binom{2i}{i}}S_1(i)
&=&
6 \zeta_2 \ln(2) + \frac{7}{2} \zeta_3 +\sqrt{\frac{\pi}{N}}\Biggl\{
4 + \frac{7}{18} \frac{1}{N} - \frac{817}{2400} \frac{1}{N^2} + \frac{3835}{37632} 
\frac{1}{N^3} + O\left(\frac{1}{N^4}\right) \Biggr\}\N\\
&&
-\frac{\ln(\overline{N})}{\sqrt{N}}\Biggl\{2 - \frac{5}{12} 
\frac{1}{N} + \frac{21}{320} \frac{1}{N^2} + \frac{223}{10752} \frac{1}{N^3} + 
O\left(\frac{1}{N^4}\right)\Biggr\}
~,
\end{eqnarray}
%---------------------------------------------------------------------------------
with $\overline{N} = N e^{\gamma_E}$ and $\gamma_E$ the Euler-Mascheroni number.
The poles in the complex plane are situated at non-positive integers and half-integers 
and (\ref{eq:TF2}) is thus a meromorphic function. The half-integer poles emerge 
algebraically and from structures like
%---------------------------------------------------------------------------------
\begin{eqnarray}
\binom{2N}{N} \frac{1}{4^N} = \frac{\Gamma(N+1/2)}{\sqrt{\pi} \Gamma(N+1)}~.
\end{eqnarray}
%---------------------------------------------------------------------------------
With these properties it can be 
defined in the complex $N$-plane. By similar arguments the analytic continuation 
for the whole $T_F^2 C_{F,A}$-contribution to $A_{gg,Q}$ is obtained. 

First results on massive OMEs with two fermion lines for $m_1 \neq m_2$ have been 
reported in \cite{Ablinger:2011pb,Ablinger:2012qj} for the moments $N = 2,4,6$ for 
$A_{Qg}$. The calculation has been performed by mapping the OMEs to tadpoles which were
computed using {\tt qexp} \cite{QEXP}.
The renormalization of these matrix elements generalizes the case of $n_f$ 
massless and one massive fermion \cite{Bierenbaum:2009mv} and is given in 
Ref.~\cite{BW13}. Since $m_c^2/m_b^2 \sim 1/10$, charm cannot be treated as massless
at the scale $m_b^2$, which makes the use of the variable flavor scheme, see 
\cite{Bierenbaum:2009mv}, built on this assumption, very problematic in case of these 
contributions. On the other hand, the fixed flavor number scheme which can be used
in precision deep-inelastic world data analyses, cf. \cite{Alekhin:2012ig}, can 
naturally 
accommodate these terms. The corresponding calculation in case of general 
values of $N$ is underway \cite{BRSW13}.
%%%%%%%%%%%%%%%%%%%%%%%%%%%%%%%%%%%%%%%%%%%%%%%%%%%%%%%%%%%%%%%%%%%%%%%%%%%%%%%%%%%%%%%%%%
\section{Ladder and Benz Topologies}
%%%%%%%%%%%%%%%%%%%%%%%%%%%%%%%%%%%%%%%%%%%%%%%%%%%%%%%%%%%%%%%%%%%%%%%%%%%%%%%%%%%%%%%%%%

\vspace{1mm}
\noindent
In the following we discuss higher topologies which contribute to the massive
OMEs $A_{ij}$ at 3--loop order. These are ladder-, Benz-, and crossed box topologies
with the respective local operator insertions (of up to four lines). The basic ladder
topologies have been calculated in Ref.~\cite{Ablinger:2012qm} up to 3-leg operator 
insertions. Here we consider the case of only one massive fermion line.
As has been described in \cite{REP} the Feynman diagrams can be represented as 
multiply nested sums.
The ladder diagrams with six massive propagators have representations in terms 
of the Appell function $F_1$ \cite{APPELL}. Most of the integrals can be solved
using {\tt Sigma} \cite{sigma}, including the pole structure in $\varepsilon$. This 
is presently more involved in case of operator insertions with more than two legs at 
six massive lines. 

For the non-divergent graphs the extension of the method \cite{Brown:2008um} to local
operators and massive lines allows the calculation. We consider the diagram shown
in Fig.~1.
%-------------------------------------------------------------------------------------------
\begin{figure}[H]
\centering
\includegraphics[scale=1.1]{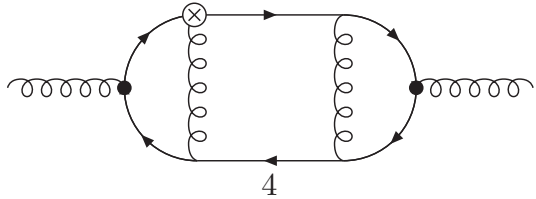}
\caption{3-loop ladder diagram containing a 3-vertex local operator insertion.}
\label{ladderA}
\end{figure}
%-------------------------------------------------------------------------------------------
The local operator insertion can be resummed introducing a subsidiary parameter $x$
%-----------------------------------------------------------------------------
\begin{eqnarray}
\sum_{j=0}^{N} T_{4a}^{N-j} T_{4b}^j &\rightarrow& \sum_{N=0}^\infty \sum_{j=0}^{N} x^N 
T_{4a}^{N-j} T_{4b}^j
= \sum_{N=0}^\infty \frac{
  (T_{4a} x)^N
- (T_{4b} x)^N}{T_{4a} - T_{4b}} \N\\ &=& \frac{1}{T_{4a} - T_{4b}} \left[
  \frac{1}{1 - xT_{4a}}
- \frac{1}{1 - xT_{4b}} \right]
= \frac{x}{(1- x T_{4a})(1- x T_{4b})}~.
\label{eq:XOP}
\end{eqnarray}
%-----------------------------------------------------------------------------
The integral may then be performed using the method \cite{Brown:2008um} and is 
expressed in terms of hyperlogarithms $\LL_{\vec{a}}(x)$. They obey the
relations
%-----------------------------------------------------------------------------
\begin{eqnarray}
\LL_{b,\vec{a}}(x) = \int_0^x \frac{dy}{y-b} \LL_{\vec{a}}(y),~~~\LL_{\emptyset}(y) =
1;~~~
\LL_{\underbrace{\mbox{\scriptsize 0, \ldots ,0}}_{\mbox{\scriptsize
$n$}}}(x)  = \frac{1}{n!} \ln^n(x),~~a_i, b \in \mathbb{R}~.
\end{eqnarray}
%-----------------------------------------------------------------------------
As an intermediary result one obtains
%-----------------------------------------------------------------------------
\begin{eqnarray}
\tilde{I}_4(x) &=& \Biggl[ - \frac{1+x}{x^3} \LL_{-1} - \frac{2x-1}{x^3} \LL_{1/2} - 
\frac{3(1-x)}{x^3}
\LL_{1} - \frac{1-2x+x^2}{(1-x)x^3} \LL_{0,-1} + \frac{1 - 2 x^2}{x^3} \LL_{0,1/2}  
\N\\ &&
-\frac{3-4x-3x^2+3x^3}{(1-x)x^3} \LL_{0,1} -\frac{1-2x^2}{x^3} \LL_{1,1/2} 
+\frac{(1-x)(2+3x)}{x^3}
\LL_{1,1} \Biggr] \zeta_3
\N\\ 
&& +\frac{(1+x) }{2 x^3} \left(3
\LL_{-1,0,0,1} - 2 \LL_{-1,0,1,1} - 3 \LL_{1,0,0,1}\right) +\frac{1}{x^2} \left(6 
\LL_{0,0,1,1} - 4
\LL_{0,1,0,1} - \LL_{0,1,1,1} \right)
\N\\
&& -\frac{(-1+2 x) }{2 x^3} \left[ 3 \LL_{1/2,0,0,1} -
\LL_{1/2,0,1,1} - 3 \LL_{1/2,1,0,1} + \LL_{1/2,1,1,1} \right] \N\\ && -\frac{3 }{2 x^2} 
\LL_{1,0,1,1}
+\frac{2 }{x^2} \LL_{1,1,0,1} -\frac{(-1+x) }{2 x^3} \LL_{1,1,1,1} +\frac{2 }{x^2} 
\left(\LL_{0,1,1} -
\LL_{1,0,1} \right) \N\\ && +\frac{\left(-1+2 x+x^2\right) }{2 (-1+x) x^3} \left[ 3 
\LL_{0,-1,0,0,1} - 2
\LL_{0,-1,0,1,1} \right] \N\\ && -\frac{5 }{-1+x} \LL_{0,0,0,1,1} -\frac{5 }{2 (-1+x)} 
\LL_{0,0,1,0,1} 
+\frac{3 (3+x) }{2 (-1+x) x} \LL_{0,0,1,1,1} \N\\ && -\frac{\left(-1+2 x^2\right) }{2 
x^3} \left[3
\LL_{0,1/2,0,0,1} + \LL_{0,1/2,0,1,1} + 3 \LL_{0,1/2,1,0,1} - \LL_{0,1/2,1,1,1} \right]
\N\\
&& +\frac{3 \left(1-3 x^2+3 x^3\right) }{2 (-1+x) x^3} \LL_{0,1,0,0,1} +\frac{8-14
x+5 x^2+3 x^3}{2 (-1+x) x^3} \LL_{0,1,0,1,1} \N\\ && +\frac{8-15 x+3 x^2}{2 (-1+x) x^3} 
\LL_{0,1,1,0,1}
-\frac{3 (-3+2 x) }{2 x^3} \LL_{0,1,1,1,1} +\frac{-6+3 x+5 x^2}{x^3} \LL_{1,0,0,1,1} 
\N\\ && +\frac{2
(-1+x)}{x^3} \LL_{1,1,1,0,1} +\frac{4-2 x+5 x^2}{2 x^3} \LL_{1,0,1,0,1} -\frac{-4+6 x+3 
x^2}{2 x^3}
\LL_{1,0,1,1,1} \N\\ && +\frac{\left(-1+2 x^2\right) }{2 x^3} \left[ 3 
\LL_{1,1/2,0,0,1} -
\LL_{1,1/2,0,1,1} \right] -\frac{3 (-1+x) (4+3 x)}{2 x^3} \LL_{1,1,0,0,1} \N\\ && 
-\frac{\left(-1+2
x^2\right)}{2 x^3}
  \left[ \LL_{1,1/2,1,0,1}
+  \LL_{1,{1}/{2},1,1,1} \right]
-\frac{(-1+x) (5+3 x) }{2 x^3} \LL_{1,1,0,1,1}~.
\label{eq:LL4}
\end{eqnarray}
%-----------------------------------------------------------------------------
Here iterated integrals over the alphabet $\{0,1,-1,1/2\}$ contribute.
Now the $N$th Taylor coefficient has to be obtained for $\tilde{I}_4(x)$, which is
possible using the package {\tt HarmonicSums} \cite{GENS2}~:
%-----------------------------------------------------------------------------
\begin{eqnarray}
\hat{I}_4(N) &=&
\frac{P_1}{2 (1+N)^5 (2+N)^5 (3+N)^5}
+\frac{P_2}{(1+N)^2 (2+N)^2 (3+N)^2} \zeta_3
\N\\&&
+\frac{(-1)^N \left(65+101 N+56 N^2+13 N^3+N^4\right) }{2
  (1+N)^2 (2+N)^2 (3+N)^2} S_{-3}
+\frac{\left(-24-5 N+2 N^2\right) }{12 (2+N)^2 (3+N)^2} S_1^3
\N\\&&
-\frac{1}{2 (1+N) (2+N) (3+N)} S_2^2
+\frac{1}{(2+N) (3+N)} S_1^2 S_2
\N\\&&
+ \frac{314+631 N+578 N^2+288 N^3+68 N^4+5 N^5}{4 (1+N)^3 (2+N)^2
    (3+N)^2} S_1^2
-\frac{3}{2} S_5
\N\\&&
-\frac{\left(399+2069 N+2774 N^2+1510 N^3+349 N^4+27 N^5\right)}{6 (1+N)^2
  (2+N)^2 (3+N)^2} S_3
-2 S_{-2,-3}
\N\\&&
-2 \zeta_3 S_{-2}
-S_{-2,1} S_{-2}
+\frac{(-1)^N \left(65+101 N+56 N^2+13
    N^3+N^4\right)}{(1+N)^2 (2+N)^2 (3+N)^2} S_{-2,1}
\N\\&&
+\frac{\left(59+42 N+6 N^2\right)}{2 (1+N)  (2+N) (3+N)} S_4
+\frac{(5+N)}{(1+N) (3+N)} \zeta_3 S_{1}\Bigl(2\Bigr)
\N\\&&
-\frac{752+2087 N+2490 N^2+1580 N^3+558 N^4+105 N^5+8 N^6}{4 (1+N)^3 (2+N)^2 
    (3+N)^2} S_2
-\zeta_3 S_2
\N\\
&&
-\frac{3}{2} S_3 S_2
-2 S_{2,1} S_2
+\frac{\left(99+225 N+190 N^2+65 N^3+7 N^4\right)}{2 (1+N)^2
  (2+N)^2 (3+N)} S_{2,1}
\N\\&&
+\frac{P_3}{(1+N)^4 (2+N)^4
    (3+N)^4} S_1
-\frac{(11+5 N) }{(1+N) (2+N) (3+N)} \zeta_3 S_1
\N\\&&
-\frac{\left(470+1075 N+996 N^2+447 N^3+96 N^4+8 N^5\right)}{4 (1+N)^2
  (2+N)^2 (3+N)^2}  S_2 S_1
-S_{2,3}
\N\\
&&
+\frac{(53+29 N)}{2 (1+N) (2+N) (3+N)}  S_3 S_1
-\frac{3 (3+2 N)}{(1+N) (2+N) (3+N)} S_1 S_{2,1}
\N\\&&
+\frac{\left(-79-40 N+N^2\right)}{2 (1+N) (2+N) (3+N)} S_{3,1}
-3 S_{4,1}
+S_{-2,1,-2}
\N\\&&
+\frac{2^{1+N} \left(-28-25 N-4 N^2+N^3\right)}{(1+N)^2 (2+N)
  (3+N)^2} S_{1,2}\left(\frac{1}{2},1\right)
-\frac{\left(-7+2 N^2\right)}{(1+N) (2+N) (3+N)} S_{2,1,1}
\N\\
&&
+5 S_{2,2,1}
+6 S_{3,1,1}
+\frac{2^N \left(-28-25 N-4 N^2+N^3\right)}{(1+N)^2 (2+N)
  (3+N)^2}  S_{1,1,1}\left(\frac{1}{2},1,1\right)
\N\\  
&&
-\frac{(5+N)}{(1+N) (3+N)} S_{1,1,2}\left(2,\frac{1}{2},1\right)
-\frac{(5+N) }{2 (1+N) (3+N)} S_{1,1,1,1}\left(2,\frac{1}{2},1,1\right)~.
\end{eqnarray}
%-----------------------------------------------------------------------------
Here $P_i$ denote polynomials, cf.~\cite{Ablinger:2012qm}.
The final expression contains individual terms which grow $\propto 2^N$ for $N 
\rightarrow \infty$. However, this singularity cancels in $\hat{I}_4(N)$. As an 
extension of the usual harmonic sums \cite{HSUM} also generalized sums 
\cite{Moch:2001zr,GENS2} contribute with weights $x_i \in \{1,1/2,2\}$.

Let us now turn to the graph shown in Fig.~2. It consists out of two contributions 
with respect to the operator insertion, cf.~\cite{Bierenbaum:2009mv}, which may be 
viewed as being obtained by contraction of the central lines of the ladder graph, resp.
the graph of the crossed box, with central operator insertion. 
%-------------------------------------------------------------------------------------------
\begin{figure}[H]
\centering
\includegraphics[scale=1.1]{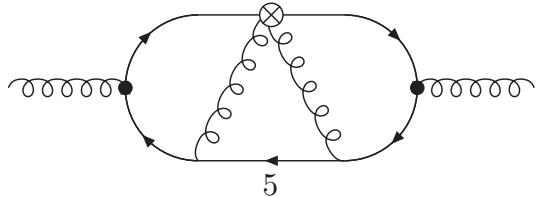}
\caption{3-loop ladder diagram containing a 4-vertex local operator insertion.}
\label{ladderB}
\end{figure}
%-------------------------------------------------------------------------------------------

\noindent
While the calculation of the former graph is rather straightforward, in the latter case
root-letters appear in the integration formalism using the method of hyperlogarithms.
Here it is possible, however, to move the corresponding root-expressions completely 
into the 
argument of the hyperlogarithms. In the next step we would like to determine the 
$N$th Taylor coefficient of the expression obtained. A first possibility consists in 
calculating a large number of Mellin-moments for the expression in an efficient way, 
which needs the use of {\tt Form} \cite{FORM} beyond the representations in {\tt Maple} 
\cite{MAPLE}. In the present case about 1500 moments have been calculated. The method
of guessing \cite{KAUERS} allows to derive a corresponding difference equation, for 
which we needed $\sim 700$ moments in the present case. This equation can now be solved
using {\tt Sigma} \cite{sigma} and the $N$th moment is obtained. In an earlier 
investigation we have determined all 3-loop anomalous dimensions and massless Wilson 
coefficients from their moments in this way, which required 5114 moments, 
\cite{Blumlein:2009tj}. Although not expected to fail with a significant probability, 
and having a large verification space with $\sim 1500$ moments available at $\sim 700$ 
needed, still the $N$th moment shall be derived having a proof certificate. A more
involved calculation using {\tt HarmonicSums} \cite{GENS2} and {\tt Sigma} \cite{sigma}
provides this. Here, difference equations of up to order {\sf o = 16} 
and degree {\sf d = 108} have to be solved. The results are presented in \cite{ABSW13}.

The emergence of root-expressions in $x$ implies in the present result quite a series 
of new sums, which are of the nested binomial- and inverse-binomial type
and also contain generalized harmonic sums. We have translated the result into 
$x$-space, where new iterated integrals with various root-type letters were obtained, 
extending those having appeared in \cite{Aglietti:2004tq}. Using the methods of
Ref.~\cite{BRON} the relative transcendence of the different functions can be checked 
and we derived a corresponding basis for these functions, cf.~\cite{ABRS13}.
In $N$-space the basis representation, including the new sums, is derived using
{\tt Sigma}, \cite{sigma}.

The method of hyperlogarithms also allows the calculation of non-singular Benz-graphs,
which have representations in terms of harmonic sums and generalized harmonic sums,
cf.~\cite{Ablinger:2012sm,ABSW13}.
%%%%%%%%%%%%%%%%%%%%%%%%%%%%%%%%%%%%%%%%%%%%%%%%%%%%%%%%%%%%%%%%%%%%%%%%%%%%%%%%%%%%%%%%%%
\section{Conclusions}
%%%%%%%%%%%%%%%%%%%%%%%%%%%%%%%%%%%%%%%%%%%%%%%%%%%%%%%%%%%%%%%%%%%%%%%%%%%%%%%%%%%%%%%%%%

\vspace{1mm}
\noindent
We reported on recent progress in calculating the asymptotic heavy flavor Wilson 
coefficients contributing to the deep-inelastic structure function $F_2(x,Q^2)$ at 
3-loop order. As a first class all contributions of $O(n_f T_F^2 C_{A,F})$ have been 
completed and basis integrals for the class $O(T_F^2 C_{A,F})$ were obtained.
Furthermore, the automatic calculation of related topology classes for other color 
factors started. We extended the former analysis to the case in which two heavy quarks 
of different mass contribute and obtained a series of Mellin moments. These 
contributions are no longer in accordance with the variable flavor number scheme, since
charm does not decouple at the mass scale of the bottom quark. In case of two massive
fermion lines with equal mass new sums and iterated integrals appear beyond the usual 
harmonic sums and polylogarithms, also leading to more singularities in the complex 
plane. We have also calculated the scalar graphs contributing to ladder-topologies
for up to six massive fermion lines, including graphs with local 4-leg operators.
Here the results are given in terms of special classes of generalized sums, which 
individually may even diverge exponentially for $N \rightarrow \infty$. This divergence 
is canceled between different terms 
contributing. In case of the graph shown in Fig.~2 a larger amount of nested binomial 
and inverse binomial sums  weighted with harmonic sums and their generalization, also 
including cyclotomic sums, emerge, extending the known alphabets both for the sums
and the associated iterated integrals. This all is invisible at the level of Mellin 
moments 
since the corresponding expressions are given in terms of rationals and single 
$\zeta$-values. It appears that in the presence of a single mass already at 3-loop 
order rich new structures are contributing in the single differential case being
characterized either by the Mellin variable $N$ or the momentum fraction $x$.

\vspace*{2mm}
\noindent
{\bf Acknowledgment.} ~We would like to thank F.~Brown for discussions. 
The Feynman diagrams have been drawn using {\tt Axodraw} \cite{Vermaseren:1994je}.
This work has  
been supported in part by DFG Sonderforschungsbereich Transregio 9, Computergest\"utzte 
Theoretische Teilchenphysik, by the Austrian Science Fund (FWF) grant P203477-N18, by 
the EU Network {\sf LHCPHENOnet} PITN-GA-2010-264564, and ERC Starting Grant PAGAP 
FP7-257638. We thank WolframResearch for their kind cooperation within {\sf 
LHCPHENOnet}. 

%%%%%%%%%%%%%%%%%%%%%%%%%%%%%%%%%%%%%%%%%%%%%%%%%%%%%%%%%%%%%%%%%%%%%%%%%%%%%%%%

\end{document}